\documentclass[letterpaper,preprint]{revtex4}
\usepackage{graphicx}
\begin{document}
\title{
On the role of coupling in mode selective excitation using
ultrafast pulse shaping in stimulated Raman spectroscopy }
\author{~S.~A.~Malinovskaya, ~P.~H.~Bucksbaum, and ~P.~R.~Berman}
\affiliation{ Michigan Center for Theoretical Physics, FOCUS
Center, and Department of Physics, University of Michigan, Ann
Arbor, MI 48109 }
\begin{abstract}
 The coherence of two, coupled two-level systems,
representing vibrational modes in a semiclassical model, is
calculated in weak and strong fields for various coupling schemes
and for different relative phases between initial state
amplitudes.  A relative phase equal to $\pi$ projects the system
into a dark state. The selective excitation of one of the two,
two-level systems is studied as a function of coupling strength
and initial phases.
\end{abstract}
\maketitle
\section{Introduction}
Investigations of coherent dynamics in multidimensional systems
induced by ultrafast shaped pulses is being carried out in a
number of research groups
\cite{Ze00,St99,Ka03,Mo02,We01,We90,Du02,Pu98,Ho00}. Special
attention is paid to the development of femtosecond laser
techniques for control over molecular motion in stimulated Raman
spectroscopy, \cite{Ka03,Or02,We01,We90,Dh94}. In these works the
selective Raman excitations are achieved with femtosecond laser
pulses with spectral or phase modulation.
The goal is to prepare a specific quantum state to facilitate
'unusual'
 structural changes or chemical reactions, \cite{Pu98,Ho00,Ri00,Sh03}. Another
fundamental application is related to the development of quantum
memory for information storage, \cite{Ch01}. A composition of
two-level systems, e.g., vibrational normal modes within an
ensemble of molecules,
  driven by an optical field, may serve as a storage device of quantum
  information.

One of the important steps needed for efficient control of
molecular motion is an understanding of the factors that govern
the system's time evolution. In this paper femtosecond pulse
shaping is discussed that allow for the selective excitation of
unresolved, coupled Raman transitions.

In \cite{We01}, a comparative analysis of Raman spectra of liquid
methanol and a mixture of benzene and deuterated benzene showed
experimental evidence for the dependence of the selective
excitation on intramolecular coupling of normal vibrational modes
in a molecule. In this paper a semiclassical model is developed
for the interaction of a shaped ultrafast laser pulse with two,
coupled two-level systems, representing coupled vibrational modes
in a single molecule. Specific questions about the mechanisms of
interaction of an external field with molecular media are
addressed such as how the coupling via an external field
influences the controllability of selective excitation and how the
result depends on the coupling strength. We also investigate the
ways of implementing a coupling mechanism for coherent control.
Within our model selective, high level coherence can be achieved
in two coupled two-level systems by a choice of the relative phase
of the initially populated states. For a particular phase, a dark
state is formed with zero eigenvalue of the interaction
Hamiltonian. It is known that for a quantum system in a dark state
the prevention of decoherence may be achieved, e.g., \cite{Be00}.
Molecules prepared in such a state may be useful for quantum
computation and communication.

\section{BASIC FORMALISM}

A semiclassical model is used to describe the interaction of an
ultrafast laser pulse with a molecular medium using stimulated
Raman scattering. The model is represented schematically in Fig.1,
where two, two-level systems describe two Raman active modes in a
molecule. Levels $|1>$ and $|3>$ are at zero relative energy,
while levels $|2>$ and $|4>$ have energies
 $\hbar \omega_{2}$ and  $\hbar \omega_4$, respectively.
Transition dipole moment matrix elements of the levels with a
virtual intermediate state $|b>$ are equal to $\mu_{ib}$.
Generally they may be different.
We investigate the effects in weak and strong fields caused by
coupling between normal vibrational modes in a molecule and by the
relative phase of the amplitudes of the initially populated
states.
Transition matrix elements of the 3-4 two-level system are assumed
to be equal, $\mu_{3b}=\mu_{4b}$, and transition matrix elements
of the 1-2 two-level system satisfy the following conditions
$\frac{\mu_{1b}}{\mu_{3b}}=\frac{\mu_{2b}}{\mu_{3b}}=r.$
The equations of motion for the probability amplitudes of two
coupled two-level systems are:
\begin{equation}
i \frac{d}{dt} \left( \begin{array}{c}
a_1 \\ a_2 \\ a_3 \\ a_4 \\ \end{array} \right)
= - \chi  \left( \begin{array}{cccc} r^2 & r^2 & r & r\\
r^2 & r^2 - \frac{\omega_2}{\chi} & r & r \\ r & r & 1 & 1\\
r & r & 1 & 1- \frac{\omega_4}{\chi} \\
\end{array} \right)  \left( \begin{array}{c}
a_1 \\ a_2 \\ a_3 \\ a_4 \\ \end{array} \right).
\end{equation}
where $\chi$ is a time-dependent variable equal to
$\frac{|\mu_{3b}|^2}{4 \hbar^2  \Delta} I(t)$, $I(t)$ is the pulse
intensity
 envelope and $\Delta$ is the
detuning of the frequency of the pulse from the frequency of the
virtual state $|b>$. Note, that in our model the pulse intensity
envelope $I(t)$ is the same for all transitions. The specific form
for the pulse shape is taken such that in the weak field regime
the pulse selectively excites transitions of predetermined
frequencies, while in strong fields the result depends on the
intensity of the field $I_0$. The intensity envelope I(t) is
defined as a real part of the Fourier transform of a function
$f(\omega)$ specified as

\begin{equation}
\tilde{I}(\omega)=I_0 \left( e^{-\alpha \omega^2}-
e^{-(\omega-\omega_4)^2 T^2} \left( 1 - e^{-(\omega-\omega_2)^2
{T_1}^2}\right) \right),
\end{equation}
where $\alpha$, $T$ and $T_1$ are free parameters. In the vicinity
of resonances this function has the spectral profile  identical to
the one suggested in \cite{Ma04} except for the fact that it
contains a "dc" component centered at $\omega=0$, to ensure that
I(t) is positive.

The real part of the inverse Fourier transform of the spectral density (2)
gives a temporal pulse function:

\begin{eqnarray}
&I(t)=I_0 C \left( 1 - A/C cos(\omega_{4}t) - B/C
cos((\omega_{2}-\Delta\omega\frac{T^2}{\tau^2}) t) \right),\\
&A=(\sqrt2 T)^{-1} e^{-\frac{t^2}{4T^2}}, B=(\sqrt2 \tau)^{-1}
e^{-\Delta \omega^2 T^2 (1-\frac{T^2}{\tau^2}) -
\frac{t^2}{4\tau^2}}, C=(\sqrt2 \tau)^{-1} e^{ -
\frac{t^2}{4\tau^2}}  \nonumber\\
&\tau^2=T^2+T_1^2, \Delta \omega= \omega_4 - \omega_2.\nonumber
\end{eqnarray}
The temporal profile of the pulse function is shown in Fig.2(a)
for parameters corresponding to experimental data on the molecular
gas $CO_2$ \cite{We01}, $T=3, T_1=3$, and $\omega_2=1,
\omega_4=1.1$, where $\omega_i$ are in frequency units, and
$T,T_1$ are in inverse frequency units. The chosen parameters $T,
T_1$ give a pulse duration corresponding to an impulsive regime of
interaction \cite{We90}. In Fig.2 (b) the Fourier transform
$\tilde{I}(\omega)$ of Eq.(3) is presented showing significantly
larger intensity at frequency $\omega_4$ than at $\omega_2$. In
weak fields the solution of Eq.(1) is
\begin{eqnarray} & a_4= i \frac{\mu_{4b}\mu_{3b}^*}{4\Delta
\hbar^2} \int_{-\infty}^{\infty} I(t) e^{i \omega_{4} t } d t = i
\frac{\mu_{4b}\mu_{3b}^*}{4\Delta \hbar^2} \tilde{I}(\omega_{4})
, \nonumber \\
&\\
& a_2= i \frac{\mu_{2b}\mu_{1b}^*}{4\Delta \hbar^2}
\int_{-\infty}^{\infty} I(t) e^{i \omega_{2} t } d t =  i
\frac{\mu_{2b}\mu_{1b}^*}{4\Delta \hbar^2} \tilde{I}(\omega_{2}).
 \nonumber
\end{eqnarray}
As a consequence, when I(t) is used in Eq.(1), the solution in a
perturbative limit leads to a weak excitation of coherence
$|\rho_{12}|$, and an efficient excitation of coherence
$|\rho_{34}|$ with the magnitude proportional to
$\tilde{I}(\omega_{4})$.

In strong fields the exact numerical solution of Eq.(1) shows that
 a choice of the field strength parameter $I_0$
provides control over excitations resulting in maximum coherence
for either $|\rho_{34}|$ or $|\rho_{12}|$. These results are in
agreement with those published in \cite{Ma04} for two uncoupled
two-level systems.

\section{Mode coupling}

A numerical solution of the time-dependent Schr\"odinger equation
 (1) describing two coupled two-level systems was obtained with I(t)
 given by Eq.(3) and two values of r, equal to $\frac{1}{2}$ and 1.
 The results reveal the importance of the relative phase between the
initial state amplitudes $a_1$ and $a_3$. This relative phase
could be established by optical pumping into state $|1>$ and using
a Raman pulse to create the $|1>-|3>$ state coherence. Coherences
$|\rho_{12}|$ and $|\rho_{34}|$ are calculated as an average over
relative phases between initially populated states $|1>$ and
$|3>$. For $r=1$, $|\rho_{12}|$ and $|\rho_{34}|$ are shown in
Fig.3(a), (bold dashed and bold solid lines, respectively), as a
function of the dimensionless intensity of the ultrafast laser
pulse. This phase averaging is equivalent to calculating
$|\rho_{34}|$ and $|\rho_{12}|$ as a sum of two contributions
resulting from two initial population distributions:
$(\rho_{11}=\frac{1}{2}, \rho_{33}=0)$ and
$(\rho_{33}=\frac{1}{2}, \rho_{11}=0)$. This approach eliminates
the role of the phase between initially populated states.
 Coherences as a function of the intensity of the field
calculated in such a way are identical to those in Fig.3(a). Also
shown in Fig.3(a), are values of $|\rho_{34}|$ and $|\rho_{12}|$,
(represented by thin solid and dashed lines, respectively), when
there is no coupling between the two, two-level systems (obtained
formally by setting $r^2=1, r=0$ in Eq. (1)). From a comparison of
the two sets of curves it is seen that phase averaged solution for
two coupled two-level systems gives much lower values of
$|\rho_{12}|$ and $|\rho_{34}|$ than that for uncoupled systems.
In weak fields the coherence $|\rho_{34}|$ prevails over
$|\rho_{12}|$ which is in agreement with Eq.(4). The coherence
$|\rho_{12}|$ increases faster than that for the case of two
uncoupled two-level systems. In strong fields $|\rho_{12}|$ and
$|\rho_{34}|$ oscillate synchronously due to nonadiabatic
interactions with the dc component of the field. The coherence
$|\rho_{12}|$ almost always is larger than $|\rho_{34}|$ owing to
its slightly smaller transition frequency.

Numerical results show that I(t) may result in significant
differences between $|\rho_{12}|$ and $|\rho_{34}|$ in a system
with different coupling constants $\mu_i$.
For $r=\frac{1}{2}$ the dependence of $|\rho_{34}|$ and
$|\rho_{12}|$ on the intensity of the field is shown in Fig.3(b)
by bold solid and bold dashed lines, respectively.
 For such coupling
constants, the probability of population transfer between
two-level systems is equal to 1/2, between states $|1>$ and $|2>$
it is equal to 1/4, and between states $|3>$ and $|4>$ to 1. As
the result, population flows from the 1-2 to the 3-4 two-level
system, maximizing coherence $|\rho_{34}|$.

For various values of intensity of the field $I_0$ the dependence
of  $|\rho_{12}|$ and  $|\rho_{34}|$
 on the relative phase
between initially populated states $|1>$ and $|3>$ has been
considered. In Fig.4(a) the case for $I_0=2.625 \pi$, $r=1$ is
represented (which gives almost equal values of $|\rho_{12}|$ and
$|\rho_{34}|$ for the phase averaged solution).
 For phases from zero to $\pi$, $|\rho_{34}| / |\rho_{12}| > 1$, and for phases
  from $\pi$ to $2\pi$, $|\rho_{34}| / |\rho_{12}| < 1$.
Consequently, using phase
 control of the initially populated states allows one to
 enhance the coherence between desired vibrational levels.
 The behavior of $|\rho_{34}| / |\rho_{12}|$ is sensitive to the intensity of field.
 For example, when
$I_0=0.5 \pi$, (which gives a phase averaged solution
$|\rho_{12}|=0.163$ and $|\rho_{34}|=0.059$ in Fig.3(a)), the
 phase dependent calculation shows that $|\rho_{12}| / |\rho_{34}| > 1$
 for all phases
 except for $\phi=\pi$, see Fig.4(b). The maximum value of coherence $|\rho^{max}_{12}|=0.4$
 at $\phi=7\pi/4$ is much
higher than that for the phase averaged solution.
When the initial relative phase $\phi$ is equal
 to $\pi$,
coherences $|\rho_{12}|$ and $|\rho_{34}|$ are equal to zero. This
is the case for any external field. Populations of all states
exhibit no time evolution. This result indicates an existence of a
dark state as can be deduced directly from the Hamiltonian in
Eq.(1). For an arbitrary value r, the necessary conditions for a
dark state are $\frac{n_3}{n_1}=r^2$ and $\phi=\pi$.

\section{Summary}
We presented a semiclassical model of the coherent control of
excitation of Raman transitions in two coupled two-level systems
using a broad-band shaped laser pulse. We analyzed the effects
caused by the coupling between four levels via a laser field. The
coupling is shown to
 cause an efficient population transfer between two-level systems.
Effects caused by the relative phase between the initially
populated states were analyzed.
Depending on the intensity of the field, the initial phase may
cause predominance of the coherence of one two-level system with
respect to another. When the relative phase between initially
populated levels $|1>$ and $|3>$ is equal to $\pi$, two, two-level
systems encounter stationary conditions regardless of the external
field strength,
implying the existence of a molecular dark state. These results
may be useful for investigation of decoherence processes caused by
the driving field.

\section*{ACKNOWLEDGMENTS}

The authors acknowledge financial support from the National
Science Foundation (No. PHY-9987916 and No. PHY-0244841) through
the Center for Frontiers in Optical Coherent and Ultrafast Science
(FOCUS) and the U. S. Office of Army Research under Grant No.
DAAD19-00-1-0412. This work was partially supported by the
National Science Foundation through a grant for the Institute for
Theoretical Atomic, Molecular, and Optical Physics at Harvard
University and Smithsonian Astrophysical Observatory.

{}


\newpage
\begin{figure}
\includegraphics[width=10cm]{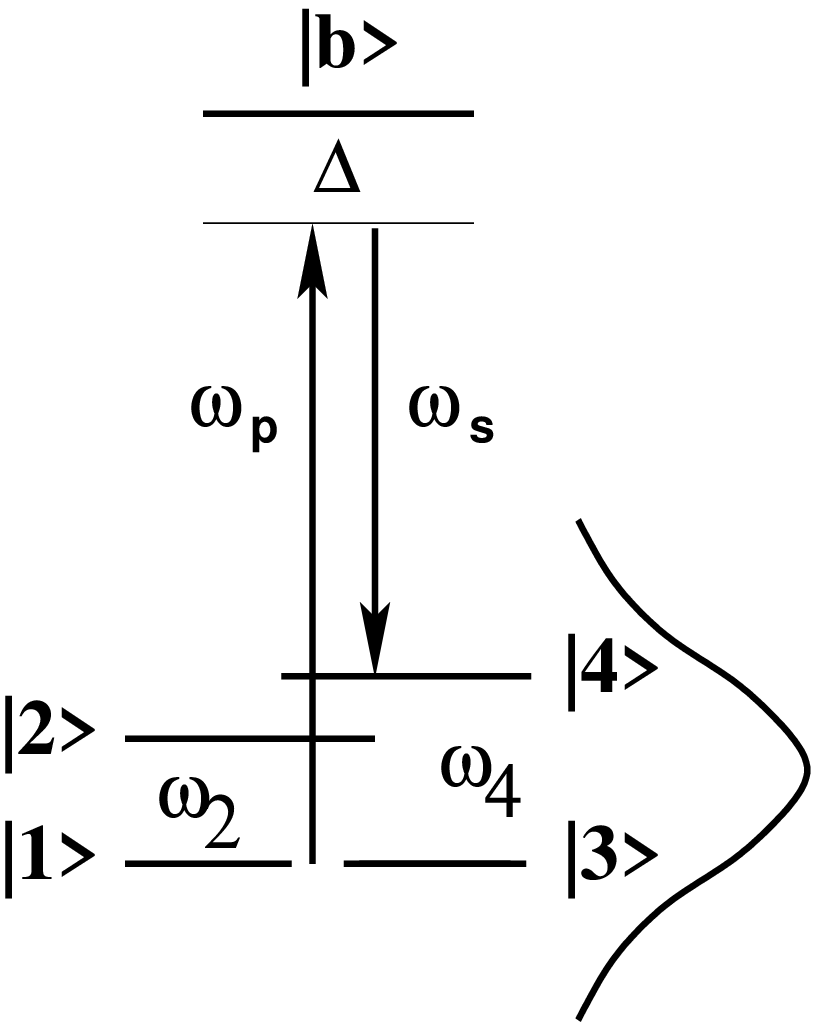}
\centerline{} \centerline{}
\caption{Schematic picture of a model system consisting of two,
two-level systems having frequencies $\omega_{2}$ and
$\omega_{4}$. Initially, the lower levels are populated evenly.
The transitions between four levels are driven by an off-resonant
femtosecond pulse.}
\end{figure}

\newpage
\begin{figure}
\includegraphics[width=10cm,angle=-0]{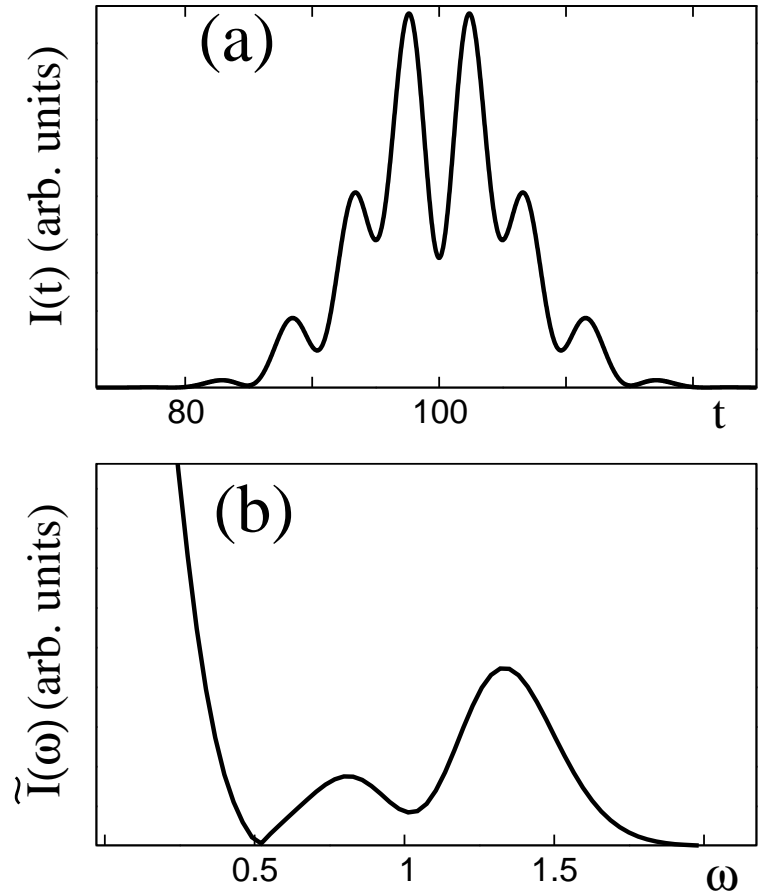}
\centerline{} \centerline{}
\caption{ (a) Intensity temporal profile I(t) for $T=3, T_1 = 3$
$[\omega^{-1}_{21}] $, (b) Fourier transform of I(t). }
\end{figure}

\newpage
\begin{figure}
\includegraphics[width=10cm,angle=-0]{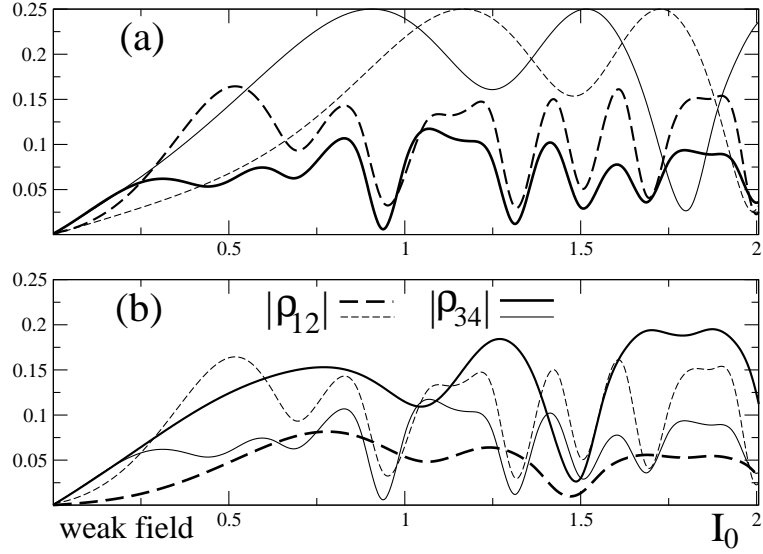}
\caption{Intensity dependence of the coherences of two, two-level
systems; $|\rho_{12}|$ is shown by dashed lines and $|\rho_{34}|$
by solid lines. In (a) bold curves depict the case for $r=1$
corresponding to two coupled two-level systems with equal coupling
constants; thin curves depict coherences for two independent
two-level systems. The phase averaged solution gives much lower
values of coherences than that for zero phase and coupling. In (b)
bold curves show coherences for $r=1/2$, and thin curves for
$r=1$. Weak coupling constants of the 1-2 system result in
efficient population flow toward the 3-4 system, strongly coupled
to the field.}

\centerline{}\centerline{}\centerline{}\centerline{}\centerline{}
\centerline{}\centerline{}\centerline{}\centerline{}\centerline{}
\centerline{}\centerline{}\centerline{}\centerline{}\centerline{}
\end{figure}
\newpage
\begin{figure}
\includegraphics[width=10cm,angle=-0]{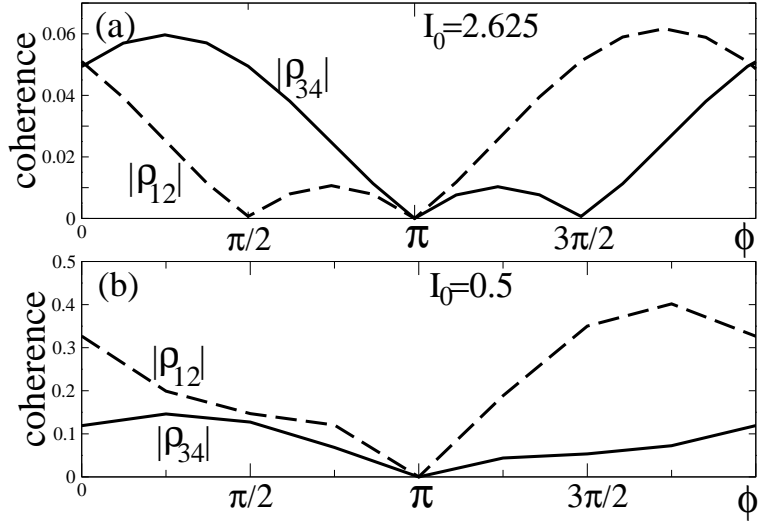}
\caption{Coherence of the 1-2 and 3-4 two-level systems as a
function of initial relative phase between levels $|1>$ and $|3>$
for $r=1$ and equal initial populations of these levels, (a)
$I_0=2.625\pi$, (b) $I_0=0.5\pi$.}
\end{figure}

\enddocument